# COHERENT CONTROL OF QUANTUM DYNAMICS WITH SEQUENCES OF UNITARY PHASE-KICK PULSES

*Luis G.C. Rego*

Departamento de Física. Universidade Federal de Santa Catarina, Florianópolis, SC, 88040-900; Brazil; E-mail: lrego@fisica.ufsc.br

*Lea F. Santos*

Department of Physics, Yeshiva University, 245 Lexington Avenue, New York, NY 10016, U.S.A.; E-mail: lsantos2@yu.edu

*Victor S. Batista*

Department of Chemistry, Yale University, P.O. Box 208107, New Haven, CT 06520-8107, U.S.A.; E-mail: victor.batista@yale.edu

ABSTRACT

Coherent optical control schemes exploit the coherence of laser pulses to change the phases of interfering dynamical pathways in order to manipulate dynamical processes. These active control methods are closely related to dynamical decoupling techniques, popularized in the field of Quantum Information. Inspired by Nuclear Magnetic Resonance (NMR) spectroscopy, dynamical decoupling methods apply sequences of unitary operations to modify the interference phenomena responsible for the system dynamics thus also belonging to the general class of coherent control techniques. This chapter reviews related developments in the fields of coherent optical control and dynamical decoupling, with emphasis on control of tunneling and decoherence in general model systems. Considering recent experimental breakthroughs in the demonstration of active control of a variety of systems, we anticipate that the reviewed coherent control scenarios and dynamical decoupling methods should raise significant experimental interest.

1. INTRODUCTION

The development of practical methods for controlling quantum dynamics with electromagnetic fields has a long history and remains an outstanding challenge of great technological interest [1-6]. In this chapter we focus on *coherent control scenarios* based on sequences of unitary pulses that can significantly influence quantum dynamics (*e.g.*, suppress tunneling) by changing the relative phases of interfering dynamical pathways, without necessarily changing the potential energy surfaces responsible for reaction dynamics in kinetic control [7, 8], or collapsing the coherent unitary evolution of the system as in control schemes based on the quantum Zeno effect [9-13]. The reviewed coherent control methods are also fundamentally different from traditional kinetic control methods where experimental conditions (*e.g.*, the effect of temperature, pressure, catalysts, or external potentials) are controlled to favor (or suppress) dynamical pathways. The coherent control sequences discussed in this review could be optimized by using closed-loop techniques where the outcome of the control process is monitored to improve the control sequence and achieve the desired dynamics [14, 15]. The discussion of such optimization methods, however, is beyond the scope of our presentation.

Quantum coherent control methods based on sequences of externally applied electromagnetic field pulses have long been considered in Nuclear Magnetic Resonance (NMR) spectroscopy. Indeed, the earliest experimental implementations of the concept of quantum coherent control (demonstrating active control over the coherent dynamics of molecular systems) date from the 1950's. Their purpose was to eliminate the undesired phase evolution by applying trains of radio-frequency (rf) $\pi$ pulses, the so-called spin-echo effect [16, 17]. During the ensuing decades the simple two-pulse technique led to a multitude of rf pulse sequences that were extensively used to study molecular structure and dynamics [18-20]. A prototype example is the WAHUHA sequence for suppression of dipolar interactions via sequences of π/2-pulses [21].

The demonstration, in 1957, that all two-level systems are mathematically equivalent suggested that coherent light pulses could lead to optical quantum control methods analogous to NMR techniques [22]. Not surprisingly, the development of the first high-power lasers in the 1960s was rapidly followed by the demonstration of the photon-echo effect [23]. However, methods to control events at the molecular scale with laser pulses were proposed only in the 1980's and bore little resemblance to NMR techniques [24, 25, 14, 26-28]. Contrary to using sequences of pulses to perturb the quantum evolution of the systems along the course of the dynamical processes, the early coherent optical control methods typically prepared the system in an initial coherent superposition by using multiple (or tailored) laser pulses. The system then evolved freely and the components of the initial coherent superposition interfered with each other while following competing relaxation pathways. Therefore, the main focus of coherent optical control methods has been the rational design, preparation and optimization of initial coherent superposition states [29-38, 15, 39, 40]. Nowadays, ultrafast lasers can produce a wide range of complex pulses with ultrashort time resolution and extremely high peak powers [41-44]. As a result, a variety of coherent control methods based on sequences of ultrafast laser pulses have been proposed [41-44], including the suppression of quantum tunneling by affecting the relative phases of interfering dynamical pathways [4-6]. The ultimate goal of these developments has been to provide fundamental understanding on how to manipulate quantum mechanical interferences in order to control the dynamics of quantum systems ranging from single atoms and molecules to quantum reaction dynamics and processes in nanoscale devices, including control over quantum superpositions of macroscopically distinct states which can be realized in Josephson junction based devices and nanomechanical resonators. To this end, decoherence is often the main obstacle hindering the achievement of coherent control. Nevertheless, coherences can be preserved even in rather complex systems as observed by photon-echo [45], pump-probe [46] and fluorescence up-conversion experiments [47] for the vibrational quantum beats of electronic states of organic dyes in liquids and condensed phases. Also, engineered solid-state devices at the nano and micro scales provide a rich ground for the observation of coherent quantum tunneling effects. Particularly, coherent charge oscillations have been produced in double quantum dot systems [48, 49] and Cooper-pair boxes (i.e., a nano-meter scale superconducting electrodes connected to a reservoir via a Josephson junction) [50]. Furthermore, in atomic physics, coherent quantum dynamics has been studied in ions and atomic Bose-Einstein condensates confined by optical traps [51].

Decoherence is a ubiquitous phenomenon in quantum system, caused by the interactions of the system with the environment. One of its consequences is the randomization of quantum phases associated with coherent superposition states, making interference and quantum control techniques ineffective. The decoherence time scales range from femtoseconds to nanoseconds for electronic excitations, due to coupling with phonons and spontaneous emission, whereas spin coherent excitations decohere in microseconds to milliseconds due to coupling to other spins in the sample. Several strategies have been proposed for suppressing decoherence, including quantum error correction schemes [52, 53] and decoherence free subspaces [54]. Here, we focus on dynamical decoupling techniques [55-58] that actively decouple the system of interest from its environment by using control pulse sequences inspired by NMR spectroscopy. The first theoretical decription of the dynamical decoupling method considered a sequence of spin echoes applied to a single spin-1/2 with the purpose of suppressing its interaction with a bosonic reservoir [56]. The method was coined quantum "bang-bang" control (after its classical analog [59-61]) referring to the ideal situation of arbitrarily strong and instantaneous pulses. The so-called 'hard pulses' in NMR are analogous to this picture. Soon afterwards, dynamical decoupling schemes were incorporated into a theoretical framework where the control operations are drawn from a discrete control group [55, 57, 58].

Quantum control methods based on dynamical decoupling have been studied in connection to a wide range of applications, including suppression of internal and external interactions, as well as control of transport behavior [62], and have become particularly popular in the area of quantum information [63, 64]. Among the various contributions to the development of dynamical decoupling methods (see Refs. [65, 66] for a more complete list of references) we mention the construction of bounded-strength Eulerian [67] and concatenated dynamical decoupling protocols [68, 69], as well as combinatorial methods for multipartite systems [70, 71]; optimized control sequences for the elimination of pure dephasing in a single qubit [72]; schemes to reduce specific decoherence mechanisms, such as 1/f noise in superconducting devices [73-76], and hyperfine- as well as phonon-induced decoherence in quantum dots [77-84]; and the compensation of imperfect averaging by adding randomized strategies into the dynamical decoupling design [85, 86, 65, 87, 66, 88, 89]. Within the field of experimental quantum information processing (QIP), dynamical decoupling techniques have found applications in liquid-state NMR [90], in solid-state systems such as nuclear quadrupole qubits [91] and fullerene qubits [92]; and have inspired charge-based [93] and flux-based [94] echo experiments in superconducting qubits beyond spin systems.

Optical control, NMR and dynamical decoupling methods share the fundamental aspect of controlling quantum dynamics by using *pulses* that affect *phases* and therefore the ensuing interference phenomena responsible for quantum dynamics. The methods have emerged from the realm of different scientific communities and continued evolving rather independently from each other for more than 30 years, partially due to the different nature of applications and the different time-scales involved. Even the concepts of *pulses* and *phases* in the different fields are often used for different physical contexts, making the connection established by common physical principles even less evident. For example, NMR and dynamical decoupling techniques applied to spin systems apply pulses that affect the phase of precession of spins (in a thermal ensemble) relative to an external field, ensuring constructive (or destructive) superposition along a desired direction in space. Therefore, the resulting effect of the pulses is to flip (or orient) the ensemble net polarization in the 3-dimensional space. Similarly, pulses of coherent optical control schemes change the phases of wavepacket components relative to the other components in a coherent superposition state and therefore rotate ket vector components in Hilbert space to ensure constructive (or destructive) interference in desired (or undesired) quantum states. As a consequence, both coherent optical control schemes and dynamical decoupling methods share common mathematical and physical principles that bridge the gap between the coherent control and the quantum information communities. It is, therefore, expected that the methods and underlying physical processes reviewed in this article should be of interest to scientific communities beyond the particular fields where the techniques were originally developed.

      The review is organized as follows. Section 2 discusses the similarities between coherent optical control and dynamical decoupling methods as contrasted to kinetic control techniques, when applied to controlling quantum dynamics in two-level model systems. Sections 3 and 4 illustrate the application of coherent optical control to manipulation of decoherence in a model quantum dot, and superexchange electron tunneling in functionalized semiconductor

nanostructures, respectively. Section 5 discusses dynamical decoupling schemes for suppression or acceleration of decoherence, for removal of unwanted internal interactions, or inducing effective couplings via external control. Section 6 presents our concluding remarks.

## 2. COHERENT OPTICAL CONTROL AND DYNAMICAL DECOUPLING

In order to illustrate the similarities between coherent optical control and dynamical decoupling methods, we consider two simple models (a) and (b) where the interference between state components in a coherent superposition is manipulated by using a sequence of unitary pulses:

(a) Particle in a symmetric double-well described by the following unperturbed Hamiltonian [5, 6]:

$$H_0(x,p) = \frac{p^2}{2} - \alpha(x^2 - \beta x^4), \quad (1)$$

with $\alpha = 1/2^2$ and $\beta = 1/2^5$. In the absence of an external perturbation, the initial non-stationary state $\Phi_0(x) = \pi^{-1/4} \exp[-(x-x_0)^2/2]$ (with $x_0 = -4$) evolves in time, tunneling back and forth through the potential energy barrier.

(b) Spin-1/2 described by the time-independent Hamiltonian,

$$H_0 = B_x \sigma_x, \quad (2)$$

where $\sigma_x, \sigma_y, \sigma_z$ are Pauli matrices and $B_x$ is a magnetic field in the $x$ direction. The initial state $\psi(0) = |\uparrow\rangle = \begin{pmatrix} 1 \\ 0 \end{pmatrix}$, an eigenstate of $\sigma_z$ evolves in time according to the coherent superposition $\psi(t) = \cos(B_x t)|\uparrow\rangle - i\sin(B_x t)|\downarrow\rangle$, oscillating between the states $|\uparrow\rangle$ and $|\downarrow\rangle$.

Both coherent dynamical processes in model systems (a) and (b) can be controlled by applying a sequence of instantaneous phase-kick pulses as described in the following sections.

2.1. COHERENT OPTICAL CONTROL OF TUNNELING: Tunneling of a particle in the double-well potential introduced by model (a) can be analyzed by considering the evolution of the non-stationary state:

$$|\Phi_0\rangle = \frac{1}{\sqrt{2}}(|\chi_0\rangle + |\chi_1\rangle), \tag{3}$$

initially localized on the left of the potential energy barrier (see Fig. 1 (a), black line), as defined in terms of the linear combination of ground and first excited states $|\chi_0\rangle$ and $|\chi_1\rangle$, with $\hat{H}|\chi_j\rangle = E_j|\chi_j\rangle$. In the absence of an external perturbation, $|\Phi_0\rangle$ evolves in time as characterized by survival amplitude,

$$\xi_t = |\langle \Phi_0 | \Phi_t \rangle|^2 = \frac{1}{2} + \frac{1}{2}\cos(\Omega t), \tag{4}$$

tunneling spontaneously back and forth through the potential energy barrier, with a Rabi frequency $\Omega = (E_1 - E_0)/\hbar$. Figure 1 (b) (black line) shows the time-dependent population $P(t) = \langle \Phi_t | h | \Phi_t \rangle$ on the right side of the potential energy barrier, where $h(x)=1$ for $x>0$ and $h(x)=0$, otherwise.

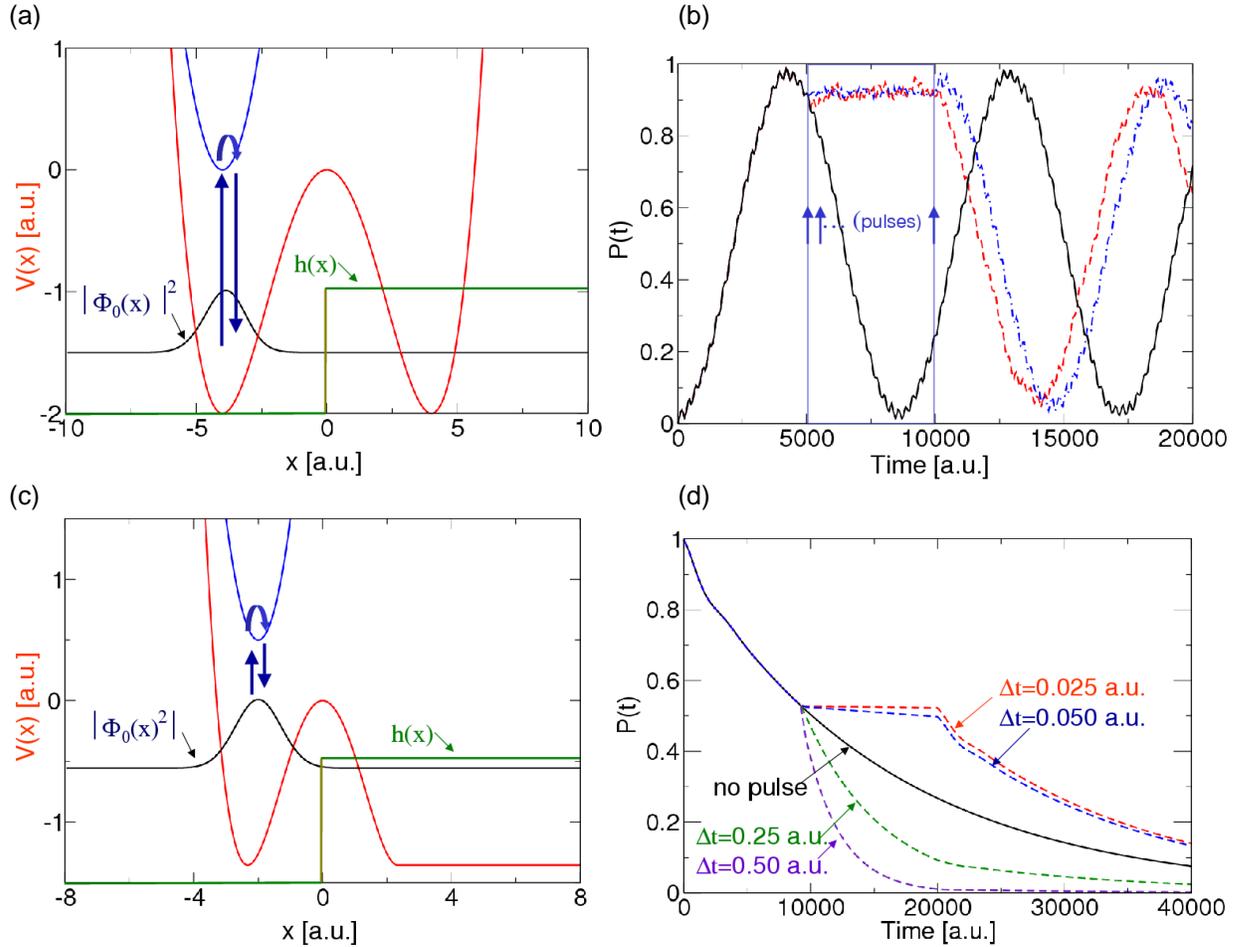

*Figure 1: Top Panels:* *(a)* Double-well model potential (red), initial state $\Phi_0(x)$ (black); and tunneling dynamics *(b)* quantified by the time-dependent population $P(t) = \langle \Phi_t | h | \Phi_t \rangle$ on the right side of the tunneling barrier in the absence of an external field (black), affected by a sequence of $2\pi$ unitary pulses (blue) stimulating resonance Raman transitions with the auxiliary state (blue arrows in panel (a)) during the time window t=5000−10000 a.u., and in the presence of a Stark perturbation (red). *Bottom Panels:* *(c)* Tunneling barrier model potential (red), initial state $\Phi_0(x)$ (black), and tunneling dynamics *(d)* quantified by the time-dependent population $P_L(t) = \langle \Phi_t | 1 - h | \Phi_t \rangle$ of the left side of the tunneling barrier in the absence of an external field (black), or influenced by a sequence of 2-π unitary pulses applied at time intervals τ=0.025 (red), 0.05 (blue), 0.25 (green), and 0.5 (purple) a.u., stimulating resonance Raman transitions with an auxiliary state (blue arrows in panel (c)).

Tunneling dynamics can be controlled in model (a) by repeatedly applying phase-kick pulses, as shown in Fig. 1 (b) (blue line) during the time window t=5000–10000 a.u. The phase-kicks are due to ultrafast laser pulses stimulating resonance Raman scattering events, as shown in Fig. 1 (a). The pulses couple $|\Phi_0\rangle$ with an auxiliary excited state $|\Phi_a\rangle$ leaving it unpopulated after and before application of the pulse (*i.e.*, $\langle\Phi_a|\Phi_t\rangle = 0$) as follows:

$$\hat{U}^{2\Theta} = \cos\left(\frac{\Gamma\tau}{2}\right)(|\Phi_0\rangle\langle\Phi_0| + |\Phi_a\rangle\langle\Phi_a|) - i\sin\left(\frac{\Gamma\tau}{2}\right)(|\Phi_0\rangle\langle\Phi_0| + |\Phi_a\rangle\langle\Phi_a|). \quad (5)$$

The results shown in Fig. 1 correspond to $2\pi$ pulses, with $\tau = 2\Theta/\Gamma$ and $\Theta = \pi$. Each pulse introduces a $\pi$ phase-shifts along the $|\Phi_0\rangle$ component of the time-evolved wavepacket $|\Phi_t\rangle$, as follows:

$$\hat{U}^{2\pi}(t) = 1 - 2|\Phi_0\rangle\langle\Phi_0|. \quad (6)$$

Considering that $N$ sufficiently frequent $2\pi$ pulses are applied at equally spaced time intervals $\Delta t = 2\tau$, starting at t_0 when $|\Phi_{t_0}\rangle = c_0(t_0)|\Phi_0\rangle + ...$ and the remaining terms in the expansion are orthogonal to $|\Phi_0\rangle$, we obtain that the time-evolved state at $t = t_0 + N\Delta t$ can be described as follows:

$$|\Phi_{2N\tau+t_0}\rangle = c_0(t_0)\left(e^{-\frac{i}{\hbar}\hat{H}\tau}\hat{U}^{2\pi}e^{-\frac{i}{\hbar}\hat{H}\tau}\right)^N|\Phi_0\rangle + ...,$$
$$= c_0(t_0)(-1)^N e^{-\frac{i}{\hbar}(E_0+E_1)2N\tau}|\Phi_0\rangle + ..., \quad (7)$$

where the second equality in Eq. (7) was obtained by substituting $\hat{U}^{2\pi}$ as defined by Eq. (6), and $|\Phi_0\rangle$ according to Eq. (3). Equation (7) shows that tunneling is completely suppressed during

the application of the phase-kick pulses since the population of $|\Phi_0\rangle$ remains constant. However, it tunneling is immediately resumed as soon as the sequence of pulses is complete.

Similar control methods based on sequences of phase-kick pulses have been applied for controlling decay and decoherence in other systems [95, 96, 56, 64, 97]. As an example, panels (c) and (d) of Fig. 1 show that a sequence of instantaneous 2-π pulses can achieve coherent control of spontaneous tunneling decay into a continuum. Here, the effect of a sequence of $2\pi$ pulses (applied during the time- window t=10000–2000 a.u. at equally spaced time intervals in the Δt=0.025–0.5 a.u. range), is quantified by the time-dependent population $P_L(t) = \langle \Phi_t | 1 - h | \Phi_t \rangle$ on the left side of the potential energy barrier. Figure 1 (c) shows that the decay of the metastable initial state can be strongly suppressed by a sequence of sufficiently frequent $2\pi$ pulses, or otherwise accelerated by less frequent sequences, without changing the potential energy surface responsible for the evolution of quantum dynamics [6]. These results are consistent with several other studies of coherent control based on sequences of $2\pi$ pulses that were successfully applied to inhibit unwanted transitions [96, 98, 99, 4-6], accelerate decay into a continuum [95], control dynamics of orientation of molecules [100, 101].

2.2. DYNAMICAL DECOUPLING OF SPIN-1/2: Dynamical decoupling methods (and, more generally, NMR techniques) aim at controlling the dynamics of a system by designing sequences of control pulses based on the desired form of the effective propagator $U$ at time $t > 0$. In general, the design of pulse sequences requires appropriate methods, the most commonly used being the average Hamiltonian theory [66] described in Sec. 5. For the particular example of model (b), the goal is to freeze the system evolution by achieving $U(t) \to 1$. It is straightforward to verify that this may be accomplished (apart from a global phase) with a sequence of instantaneous π-pulses applied perpendicularly to the $x$ direction, such as $P_z = \exp[-i\pi\sigma_z/2]$, which rotates the spin by $180^o$ around the $z$ direction. The pulses are applied after every time interval $\Delta t$ of free evolution, so that at $2\Delta t$, we obtain:

$$\begin{aligned}U(2\Delta t) &= P\exp[-iB_x\sigma_x\Delta t]P\exp[-iB_x\sigma_x\Delta t], \\ &= (-1)\exp[-iB_x\exp(i\pi\sigma_z/2)\sigma_x\exp(-i\pi\sigma_z/2)\Delta t]\exp[-iB_x\sigma_x\Delta t], \\ &= (-1)\exp[iB_x\sigma_x\Delta t]\exp[-iB_x\sigma_x\Delta t] = -1.\end{aligned} \quad (8)$$

The effect of pulsing is to reverse the system evolution, canceling out dephasing at times $t = n2\Delta t$, with $n \in N$. Equivalently to the tunneling problem discussed in Sec. 2.1, the effects of the pulses may also be understood from the perspective of the state of the system. The pulse introduces a phase change to the component $|\downarrow\rangle$ of the coherent superposition, leading to destructive interference and the subsequent restoration of the initial state (apart from a global phase), as follows:

$$\begin{aligned}\psi(0) &= |\uparrow\rangle, \\ U(\Delta t)\psi(0) &= \cos(B_x\Delta t)|\uparrow\rangle - i\sin(B_x\Delta t)|\downarrow\rangle, \\ PU(\Delta t)\psi(0) &= (-i)\left[\cos(B_x\Delta t)|\uparrow\rangle + i\sin(B_x\Delta t)|\downarrow\rangle\right], \\ U(\Delta t)PU(\Delta t)\psi(0) &= (-i)|\uparrow\rangle, \\ PU(\Delta t)PU(\Delta t)\psi(0) &= -|\uparrow\rangle.\end{aligned} \quad (9)$$

Section 5 generalizes the dynamical decoupling scheme briefly introduced in this section with more complex pulse sequences designed to eliminate non-commuting terms of the Hamiltonian. In the language of optical control, these sequences address scenarios with multiple interfering quantum paths available.

2.3. KINETIC CONTROL: Control of quantum tunneling dynamics can also be achieved by using kinetic control methods, where external electromagnetic fields are applied to affect the ensuing quantum dynamics by modulating the potential energy landscape, collapse the coherent evolution of the system, or induce mode-selective excitation. In order to compare coherent control and dynamical decoupling methods, we consider the Stark perturbation,

$$H_1(x,t) = \lambda x \sin(\omega t). \qquad (10)$$

modulating the potential energy landscape of the symmetric double-well potential (model system (a)). The parameters of the perturbation $H_1$ are chosen with a suitable resonant frequency $\omega = 0.01$ and coupling parameter $\lambda=0.003171$ [8]. Figure 1 (b) (red line) shows that such a time-dependent perturbation, applied during t=5000─10000 a.u. inhibits tunneling.

Other kinetic control methods, such as scenarios based on the quantum Zeno effect [10], collapse the coherent quantum evolution of the system due to an external perturbation [12, 13]. The perturbation can either delay (Zeno effect) or accelerate (anti-Zeno effect) the decoherence process [102]. Unifying approaches based on an adiabatic theorem [9], or considering the quantum measurement theory in detail [11], have been proposed to explain the various forms of Zeno effects produced by non-unitary pulses (measurements), unitary pulses (dynamical decoupling) or continuous strong coupling.

All of the methods discussed in this section belong to the general class of active control scenarios where the properties of engineered electromagnetic fields are externally manipulated to produce a desired outcome of a dynamical process. The distinctive aspect of coherent control and dynamical decoupling methods is that they can affect the *phases* of wavepacket components responsible for intereference, without changing the potential energy surface or collapsing the coherent evolution of dynamics. In contrast, kinetic control methods for dynamical localization, or coherent destruction of quantum tunneling, affect the potential energy surface where the system propagates [7, 8].

Many experimental studies have demonstrated the feasibility of kinetic control of quantum dynamics by applying sinusoidal driving potentials. Starting with the use of radio-frequency (rf) electromagnetic pulse sequences in nuclear magnetic resonance (NMR), rf-pulse techniques were subsequently applied to achieve coherent control in a wide range of systems, including applications to the renormalization of Landé $g$ factors in atoms [103], the micromotion of single trapped ions [104], motion of electrons in semiconductor superlattices [105], resonance activation of Brownian particle out of a potential well modeling a current-biased Josephson tunnel junction in its zero-voltage state [1] (also analyzed by theoretical studies [106-111]) and dynamical suppression of interwell tunneling of a Bose-Einstein condensate (BEC) in a strongly driven periodic optical potential [112, 113]. Several other experiments have also reported tunneling suppression [114, 115], and recently dynamical localization and coherent suppression of tunneling have been demonstrated for light propagating in coupled waveguide arrays [116, 117]. These experimental breakthroughs in the manipulation of oscillatory fields to achieve kinetic control suggest that coherent control scenarios with similar capabilities should be useful to control the quantum dynamics of systems ranging from single atoms and molecules to BEC's, nanoscale devices, including quantum-dots (QD) and quantum-dot molecules, and ultimately the control over quantum superpositions of macroscopically distinct states which can be realized in Josephson junction based devices. Therefore, the reviewed coherent control scenarios and dynamical decoupling methods should raise significant experimental interest, particularly in studies of coherent optical manipulation of electronic excitations in devices where performance is limited by quantum tunneling and decoherence.

## 3. COHERENT CONTROL OF DECOHERENCE IN A MODEL QUANTUM DOT.

Coherent control scenarios based on sequences of unitary phase-kick pulses have been recently investigated as applied to controlling decoherence in an electronic quantum dot (QD) coupled to a free-standing quasi two-dimensional silicon phonon cavity (see Fig. 2) [6]. The model allows one to investigate coherent control in a system analogous to suspended heterostructures typically built with nanomachining technology [118] [119] that exhibits rich quantum chaotic behavior [120, 121]. As an example of such heterostructures, Fig. 2 shows a square 1×1 $\mu m$ free-standing phonon cavity (50 $nm$ thick), produced by the Cornell Nanofabrication Facility, with a QD of diameter 100–250 $nm$ produced by doping selectively a circular area at the surface of the silicon plate, or by suspending metallic gates [122, 123]. The results presented in this section correspond to a QD of radius $R=125$ $nm$, placed slightly off-center in the phonon cavity at the non-symmetric position $(x, y) = (0.650, 0.575)$ $\mu m$. The position of the quantum dot is important since it determines the nature of the underlying relaxation dynamics, due to the interplay between the symmetries of the circular QD states and the square cavity of phonon modes. In particular, the spectrum of energy level spacing is regular (*i.e.*, described by a Poissonian distribution) when the QD is placed at the center of the cavity (x, y) = (0.5, 0.5) $\mu m$ [120, 121]. However, it exhibits a distinct quantum chaotic behavior characterized by a Gaussian Unitary Ensemble (GUE) of random matrices when the QD is placed at a non-symmetric position such as $(x, y) = (0.650, 0.575)$ $\mu m$ [120, 121].

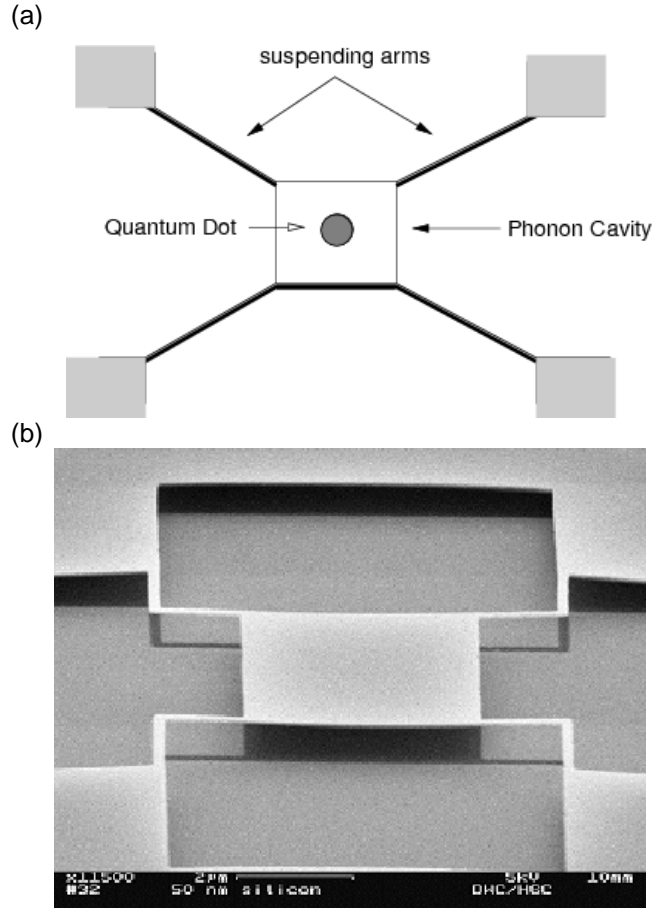

*Figure 2:* (a) Model quantum dot structure in a free-standing square phonon cavity. b) Free standing silicon plate, 50 nm thick and 4 μm long, produced by the Cornell Nanofabrication Facility.

The model Hamiltonian of the QD coupled to the phonon cavity has been described as a sum of electron, phonon and electron-phonon interactions as follows:

$$\hat{H} = \hat{H}_{el} + \hat{H}_{ph} + \hat{H}_{el-ph},$$
$$= \sum_{k} E_k b_k^{+} b_k + \sum_{\alpha}\left(\hat{n}_\alpha + \frac{1}{2}\right)\hbar\omega_\alpha + \sum_{k'\alpha k} V_{k'\alpha k} b_{k'}^{+}\left[a_\alpha^{+} + a_\alpha\right]b_{k'},$$
(11)

where the electronic basis states $|k\rangle = |l, v\rangle$ correspond to the possible states of an electron in a 2-dimensional circular quantum dot, where $l$ is the angular momentum and $v$ is the radial quantum number. The operators $a_\alpha^+$ and $a_\alpha$ create and annihilate phonon modes $\alpha$ and define the number operator $\hat{n}_\alpha = a_\alpha^+ a_\alpha$. The electron-phonon coupling terms $V_{k'\alpha k}$ depend on the material properties as well as the geometry of the structure, carrying information on the symmetry of the nanoelectromechanical system [120, 121]. The Hamiltonian of the compound QD-phonon system, therefore, can be written in the basis set of direct products of the one-electron states $|k\rangle$ and the multi-phonon states $|n_1 n_2 n_3 ... n_N\rangle$, where $n_\alpha = 0, 1, ..., n$, denotes the number of phonon quanta in mode α, considering a total of $N = 27$ distinct phonon modes and $n_\alpha \leq 30$. A typical state for the compound system can be represented as follows:

$$|k;\mathbf{n}\rangle = |k\rangle \prod_\alpha^N \frac{1}{\sqrt{n_\alpha!}} \left(a_\alpha^+\right)^{n_\alpha} |0\rangle. \tag{12}$$

The dynamics of decoherence has been investigated by computing the time evolution of the electronic angular momentum $L_{el} = Tr\{\hat{\rho}_{el}(t)\hat{L}\}$, where $\hat{\rho}_{el}(t) = Tr_{ph}\{\hat{\rho}(t)\}$ is the reduced electronic density matrix and $Tr_{ph}$ designates the trace over phonon states. The decoherence dynamics was quantified by computing $\Gamma_{el} = Tr\{\hat{\rho}^2{}_{el}(t)\}$. These calculations required the integration of the time-dependent Schrödinger equation, after diagonalization of the compound QD-phonon Hamiltonian.

The dynamics of decoherence has been manipulated by a coherent control scenario comprising a sequence of 2π pulses [6]. Each pulse has been described by the unitary operator,

$$\begin{aligned}\hat{U}^{2\pi} &= \left(\hat{U}^{2\pi}\right)_{el} \otimes \mathbf{I}_{ph}, \\ &= \left(\mathbf{I} - 2|l=1,v\rangle\langle l=1,v|\right)_{el} \otimes \mathbf{I}_{ph},\end{aligned} \tag{13}$$

and introduces a π phase-shift in the *l=1* component of the time-evolved state, with $\mathbf{I}_{ph}$ the identity matrix in the basis set of phonon states.

The initial electronic state was defined as the first excited rotational state, as defined by the undisturbed electronic states of the circular 2-dimensional QD of radius *R*,

$$\phi_k(r,\theta) = \frac{J_{|l|}\left(\alpha_{l\nu}\frac{r}{R}\right)\exp[il\theta]}{\sqrt{\pi}R\left|J_{|l|+1}(\alpha_{l\nu})\right|}, \qquad (14)$$

where $k \equiv (l,\nu)$ and $l = 0, \pm 1, \pm 2,...$, with $\alpha_{l\nu}$ the υ–th root of the Bessel function of order $|l|$, $J_{|l|}(\alpha_{l\nu}x)$. The corresponding energies of the one-electron states $|\phi_k\rangle$ are $E_k = \frac{\hbar^2}{2m_e}\frac{\alpha_{l\nu}^2}{R^2}$, with $m_e$ the electron effective mass. The initial state of the phonon bath was defined according to the density matrix, at finite temperature *T*,

$$\rho^{ph}_{nn} = \frac{\exp(-E_{ph}(\mathbf{n})/k_BT)}{\sum_{\{\mathbf{n}\}}\exp(-E_{ph}(\mathbf{n})/k_BT)}, \qquad (15)$$

where $E_{ph}(\mathbf{n}) = \sum_\alpha \left(n_\alpha + \frac{1}{2}\right)\hbar\omega_\alpha$ is the energy of the multimode phonon cavity state $\mathbf{n} \equiv (n_1, n_2,..., n_N)$.

Figure 3 (black line) shows the evolution of the time-dependent electronic angular momentum, $L_{el} = Tr\{\hat{\rho}_{el}(t)\hat{L}\}$ (panel A), and the decoherence measure $\Gamma_{el} = Tr\{\hat{\rho}^2{}_{el}(t)\}$ (panel B) during the early time relaxation after initializing the electronic state in the first excited rotational state $L_{el}=1$, with $|k\rangle = |l=1, \nu=1\rangle$ in interaction with the phonon bath at $T=200$ mK. Figure 3 (red line) shows the evolution of $L_{el}$ and $\Gamma_{el}$, corresponding to the dynamics of the system perturbed by a sequence of $2\pi$ pulses, applied at intervals $\tau = 0.9$ ns during the time-window t=0.1−1.5μs. It is shown that the decoherence dynamics is inhibited and ultimately halted by the sequence of phase-kick pulses, without collapsing the evolution of the system. Once the sequence of perturbational pulses is complete, at t=1.5μs, the decoherence dynamics is reestablished.

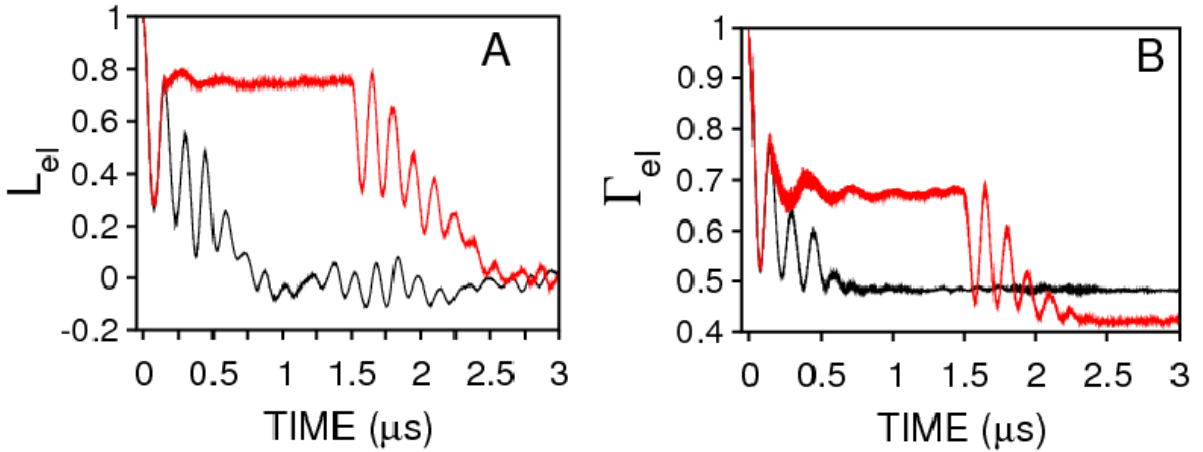

*Figure 3:* (A) Time-dependent angular momentum $L_{el} = Tr\{\hat{\rho}_{el}(t)\hat{L}\}$, and (B) decoherence measure $\Gamma_{el} = Tr\{\hat{\rho}^2{}_{el}(t)\}$ associated with the dynamics of an electron in a quantum dot structure, coupled to a 2-dimensional free-standing thermal phonon cavity. The freely evolving propagation (black line) is compared to the dynamics of the system perturbed by a sequence of $2\pi$ pulses, applied at intervals $\Delta\tau$ =0.9 ns during the time-window t=0.1-1.5μs (red line).

The results reviewed in this section show that sequences of phase-kick pulses can be applied to coherently control electronic decoherence in quantum dots coupled to a thermal bath. Considering the possibility of engineering this type of semiconductor devices where quantum tunneling and decoherence phenomena can be tested, it is natural to anticipate considerable experimental interest in examining the proposed coherent control scenario. In particular, quantum dots (QDs) have already been recognized as physical realizations of artificial atoms and molecules whose properties (*e.g.*, structural and transport) can be engineered for specific applications and modulated in the presence of external fields [124-126]. Proposals include arrays of coupled QDs for applications to create charge or spin qubit gates [48, 49, 127], or quantum memory units [128]. However, efficient methods for coherent-optical manipulation of decoherence and quantum tunneling dynamics have yet to be established.

## 4. COHERENT CONTROL OF SUPEREXCHANGE ELECTRON TRANSFER.

Recent theoretical studies have addressed the feasibility of creating and coherently manipulating electronic excitations in $TiO_2$ semiconductor surfaces, functionalized with molecular adsorbates [4, 6]. These studies aimed to explore realistic models of molecular qubits based on existing semiconductor materials, building upon previous work focused on the characterization of time-scales and mechanisms of interfacial electron transfer in sensitized $TiO_2$-anatase nanoparticles [129-132], and earlier studies of coherent optical control of molecular processes [29-31, 37].

Functionalization results from the adsorption of molecules onto the semiconductor surface. As a result, molecules are covalently attached forming surface complexes that introduce electronic states in the semiconductor band gap (see Fig. 4 (a)). The host semiconductor material is thus sensitized to photoabsorption at lower frequencies, characteristic of the molecular adsorbates, leading to ultrafast interfacial electron injection when there is suitable energy match between the photoexcited electronic states in the surface complex and the electronic states in the conduction band of the semiconductor surface. The resulting photoexcitation and interfacial relaxation process has already raised significant experimental interest since it is central in applications to photovoltaic devices for solar-energy conversion [133, 134] and photocatalysis [129, 135-139].

Recent computational studies have addressed the relaxation dynamics of electron holes left within the semiconductor band gap after photoinduced electron injection (see Fig. 4) [4, 131, 5]. The distinctive aspect of holes localized in these intraband states is that they remain off-resonance relative to the semiconductor (valence and conduction) bands, naturally protected from dissipation into the semiconductor material. However, superexchange hole-tunneling into near resonant states localized in adjacent adsorbate molecules often occurs, even under low surface coverage conditions, when the electronic states of the adsorbates are only indirectly coupled by the common host-substrate (see Fig. 4).

Computations of transient hole populations have been based on mixed quantum-classical simulations of dynamics, treating the evolution of electronic states fully quantum mechanically in conjunction with the classical propagation of an ensemble of nuclear trajectories evolving on effective mean-field potential energy surfaces [130, 4, 131, 5].

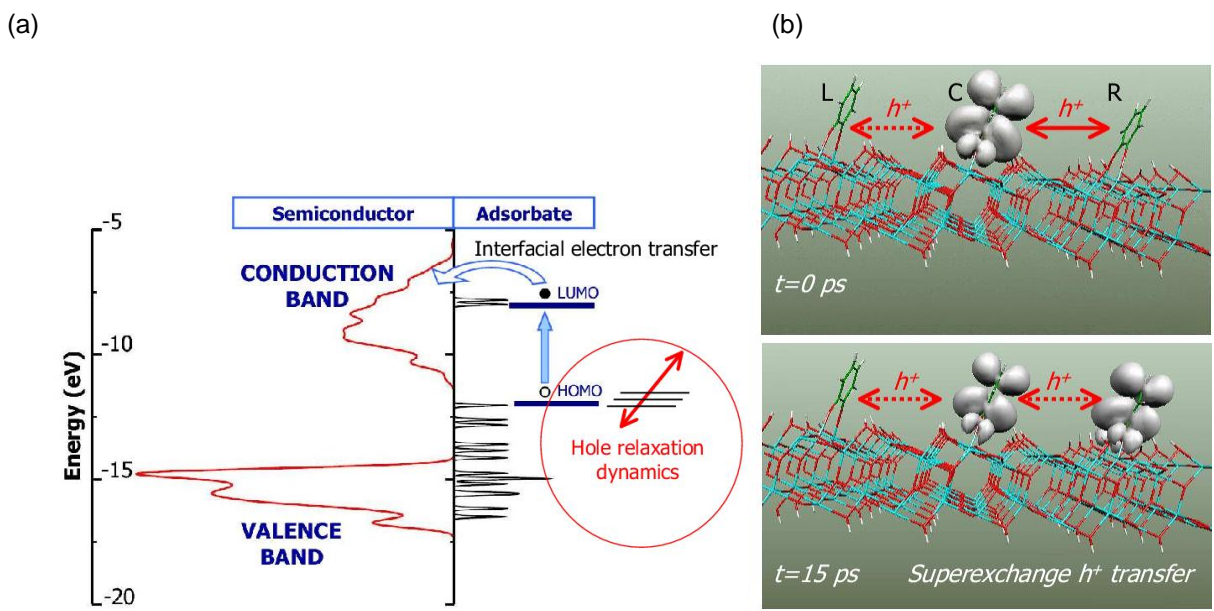

*Figure 4:* (a) Schematic energy diagram of the electronic structure of $TiO_2$-anatase surface functionalized with molecular adsorbates, including the valence and conduction bands of $TiO_2$ and the energy levels due to the molecular adsorbate. The arrows indicate the photoinjection process and the relaxation of the hole in the manifold of near-resonant energy levels localized in the adsorbates. (b) $TiO_2$-anatase functionalized with catechol molecules and isosurface density representing a nonstationary hole state delocalized on the molecular adsorbates after 15 ps of relaxation dynamics.

Figure 5 shows the evolution of time-dependent hole populations $P_{MOL}(t)$ of the three adsorbate molecules functionalizing the $TiO_2$ nanostructure shown in Fig. 4, as quantified by the diagonal elements of the reduced density matrix $\rho(t)$ associated with the subspace of electronic states localized in the adsorbate molecules MOL=(**L**eft; **C**enter; **R**ight). These results indicate that the hole tunnels between adjacent adsorbate molecules, covalently attached to the $TiO_2$ approximately 1 nm apart from each other and therefore with negligible overlap of molecular orbitals. Electronic couplings with off-resonant states in the common host substrate, however, induce superexchange hole tunneling, keeping the hole localized in the monolayer of adsorbates rather than injected into the semiconductor host substrate) [131]. The analysis of individual members of the ensemble indicates that population exchange is most prominent when the electronic states of adjacent adsorbates become near resonant (see population exchange at about 420 fs), leading to net population transfer as characterized by the Rabi oscillations shown in Fig. 5. Quantum coherences during the entire simulation time as characterized by the analysis of off-diagonal elements of the reduced density matrix (Fig. 5) and the measure of decoherence $Tr[\rho^2(t)]$ [131]. These results suggest that the observation of Rabi oscillations, associated with the adsorbate electronic populations, could provide a simple experimental probe of the predicted quantum coherent relaxation dynamics.

Computational studies have addressed the nontrivial question as to whether the underlying superexchange hole-tunneling dynamics, associated with electronic relaxation in monolayers of adsorbate molecules, could be coherently controlled by the application of (deterministic and stochastic) sequences of unitary phase-kick pulses (see Fig. 5) [4, 6]. As an example, Fig. 5 shows the perturbational effect of a sequence of $2\pi$ pulses on the relaxation dynamics of electron holes undergoing superexchange hole transfer between adsorbate molecules functionalizing a $TiO_2$ nanoparticle. The pulses are applied during the t=15—60 ps time window (indicated with arrows in Fig. 5) at intervals of 550 fs, starting at t=15 ps when there is maximum entanglement between adsorbates C and R (*i.e.*, when the off-diagonal elements $\langle C|\rho(t)|R\rangle$ are maximum). It is shown that a sequence of $2\pi$ optical pulses, resonant with electronic transitions of adsorbate C, strongly suppresses the Rabi oscillations due to superexchange hole tunneling keeping constant the hole population in adsorbate C for as long as the sequence of pulses is applied (Fig. 5, top left). The analysis of off-diagonal elements of the reduced density matrix $\rho(t)$, indicates that the resulting effect of the sequence of pulses is to affect the interference between electronic states by rapidly affecting the relative phase of states responsible for relaxation, without collapsing the coherent quantum evolution of the hole. The coherent hole tunneling is reestablished, once the sequence of phase-kick pulses is complete.

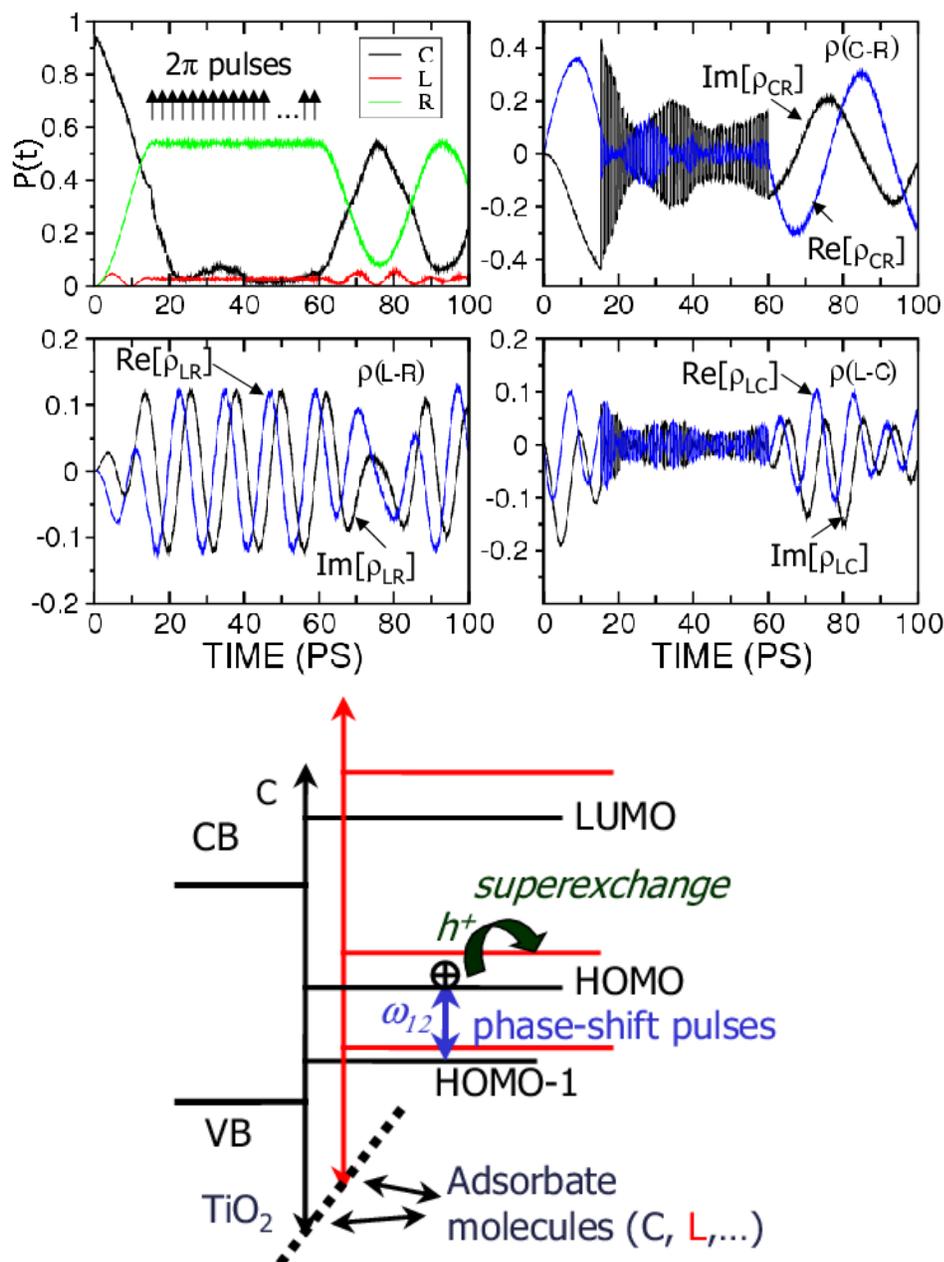

*Figure 5:* Upped panels: Time-dependent hole population P(t) for the three adsorbates C, L and R. The arrows indicate the start and the end of the sequence of $2\pi$ pulses. Other top panels show the real (black) and imaginary (blue) parts of the off-diagonal elements of the reduced density matrix indicated by the labels. Lower Panel: Schematic energy diagram of adsorbate complexes C and L, and electronic transitions associated with coherent control of superexchange hole tunneling based on multiple phase-kick pulses.

While the results illustrated in Fig. 5 correspond to deterministic sequences of $2\pi$ pulses, similar coherent control over relaxation dynamics can be achieved with stochastic $2\Theta$ pulses, where $\Theta$ is a random phase. The intervals between pulses can also be varied stochastically so long as the pulses are applied sufficiently frequently [5]. These results suggest the feasibility of applying currently available femtosecond laser technology to achieve coherent optical manipulation of electronic excitations in functionalized $TiO_2$ surfaces, under a wide range of experimental conditions.

## 5. DYNAMICAL DECOUPLING

This section reviews the basic ideas of dynamical decoupling including the underlying group theoretic framework (Sec. 5.1); strategies to improve protocol performance (Sec. 5.2), and a brief discussion of relevant frames and the control settings (Sec. 5.3). Throughout, spin-1/2 refers to a model for any two-level system, being therefore frequently exchanged by the more general idea of a qubit. Two scenarios are presented as illustrative examples: in Sec. 5.4 the evolution of an isolated spin-1/2 chain is frozen by removing unwanted internal interaction – emphasis is given to the advantages of randomization; and Sec. 5.6 discusses the suppression of decoherence in the case of a single spin-1/2 coupled to a bosonic bath - the phenomena of decoherence acceleration and asymptotic saturation are addressed.

### 5.1. THEORETICAL FRAMEWORK

In dynamical decoupling methods, a time-dependent control Hamiltonian $H_c(t)$ is added to the Hamiltonian $H_0(t)$ of the system whose dynamics we want to modify. The time evolution operator in the physical (Schrödinger) frame under the total Hamiltonian becomes

$$U(t) = \mathcal{T} \exp\left[-i\int_0^t [H_0(u) + H_c(u)]du\right], \qquad (16)$$

while the control propagator is $U_c(t) = \mathcal{T}\exp\left[-i\int_0^t H_c(u)du\right]$, where we set $\hbar = 1$ and $\mathcal{T}$ denotes time ordering. In the ideal case of bang-bang control, the pulses $P_k$ are instantaneous and depend only on $H_c(t)$, whereas during the intervals $\Delta t = t_k - t_{k-1}$ between control operations the system evolves freely according to $H_0(t)$. The propagator at $t_n = n\Delta t$, $n \in N$, is then

$$\begin{aligned}U(t_n) &= P_n U(t_n, t_{n-1}) P_{n-1} U(t_{n-1}, t_{n-2}) \ldots P_1 U(t_1, 0) P_0 \\ &= \underbrace{(P_n P_{n-1} \ldots P_1 P_0)}_{U_C(t_n)} \underbrace{(P_{n-1} \ldots P_1 P_0)^+ U(t_n, t_{n-1})(P_{n-1} \ldots P_1 P_0) \ldots (P_1 P_0)^+ U(t_2, t_1)(P_1 P_0) P_0^+ U(t_1, 0) P_0}_{\tilde{U}(t_n)},\end{aligned} \qquad (17)$$

Above, $\tilde{U}(t_n)$ stands for the evolution operator in the logical frame:

$$\tilde{U}(t) = \mathcal{T}\exp\left[-i\int_0^t \tilde{H}(u)du\right], \text{ and } \tilde{H}_0(t) = U_c^+(t) H_0(t) U_c(t). \qquad (18)$$

The logical (also known as toggling) frame corresponds to a time-dependent interaction representation that follows the control. It is a theoretical tool often used in the design of dynamical decoupling protocols along with the average Hamiltonian theory [18, 19]. The latter consists in writing the logical propagator in terms of a single exponential and identifying the appropriate sequence of pulses leading to the desired form of the effective propagator at a final time $t_n$. For a time-independent system Hamiltonian $H_0$, we find

$$\tilde{U}(t_n) = \exp[-i\tilde{H}_n \Delta t] \ldots \exp[-i\tilde{H}_2 \Delta t]\exp[-i\tilde{H}_1 \Delta t] = \exp\left[-i\sum_{k=0}^\infty \left(\overline{H}^{(k)}(t_n)\right) t_n\right], \qquad (19)$$

where $\tilde{H}_n = (P_{n-1}...P_1P_0)^+ H_0 (P_{n-1}...P_1P_0)$ are transformed Hamiltonians and the Baker-Campbell-Hausdorff expansion was used to derive the last equality [the Magnus expansion [18, 19]   is required when dealing with a time-dependent system Hamiltonian]. Each $\overline{H}^{(k)}(t_n)$ is proportional to $(\Delta t)^k / t_n$ and involves $k$ time-ordered commutators of transformed Hamiltonians.

For cyclic control with cycle time $T_c$, that is, $H_c(t + nT_c) = H_c(t)$ and $U_c(t + nT_c) = U_c(t)$, physical and logical frame coincide at every $T_n = nT_c$, therefore $U(nT_c) = \tilde{U}(nT_c)$. At these instants, the system appears to evolve under a time-independent average Hamiltonian $\overline{H} = \sum_{k=0}^{\infty} \overline{H}^{(k)}$ leading to $\tilde{U}(nT_c) = \tilde{U}(T_c)^n = \exp[-i\overline{H}nT_c]$, so that to analyze the system evolution at $T_n$ it suffices to derive the propagator at $T_c$. Pulse sequences are then constructed based primarily on the appropriate form of the dominant terms in the average Hamiltonian. Given $T_c = M\Delta t$, where $M$ is a number determined by the sequence considered, the three first dominant terms are:

$$\overline{H}^{(0)} = \frac{\Delta t}{T_c} \sum_{k=1}^{M} \tilde{H}_k, \tag{20}$$

$$\overline{H}^{(1)} = -i \frac{(\Delta t)^2}{2T_c} \sum_{l=2}^{M} \sum_{k=1}^{l-1} [\tilde{H}_l, \tilde{H}_k], \tag{21}$$

$$\overline{H}^{(2)} = -\frac{(\Delta t)^3}{6T_c} \left\{ \sum_{m=3}^{M} \sum_{l=2}^{m-1} \sum_{k=1}^{l-1} \{[\tilde{H}_m, [\tilde{H}_l, \tilde{H}_k]] + [[\tilde{H}_m, \tilde{H}_l], \tilde{H}_k]\} \right. \\ \left. + \frac{1}{2} \sum_{l=2}^{M} \sum_{k=1}^{l-1} \{[\tilde{H}_l, [\tilde{H}_l, \tilde{H}_k]] + [[\tilde{H}_l, \tilde{H}_k], \tilde{H}_k]\} \right\}. \tag{22}$$

For short times and in the limit $T_c \to 0$, reshaping the Hamiltonian based on the dominant terms leads to dynamics close enough to the desired one.

In NMR spectroscopy, the design of control protocols usually aims at very specific systems. A more general approach was developed by invoking group theory [55, 57], where the purpose is to map the dominant term $\overline{H}^{(0)}$ into a group-theoretic average $\overline{H}_G$. The pulses are successively drawn from a discrete dynamical decoupling group $G = \{g_j\}, j = 0, 1, ..., |G|-1$, where $|G|$ is the size of the group and $T_c = |G|\Delta t$, so that

$$\tilde{U}(T_c) = \prod_{j=0}^{|G|-1} U_{j+1}, \tag{23}$$

with $U_{j+1} = g_j^+ U(t_{j+1}, t_j) g_j$, $\tilde{H}_{j+1} = g_j^+ H_0 g_j$, $P_{j+1} = g_{j+1} g_j^+$, $P_0 = g_0$,

and

$$\overline{H}_G = \frac{1}{|G|} \sum_{k=1}^{|G|} \tilde{H}_k = \overline{H}^{(0)}. \tag{24}$$

In order to freeze the system evolution and achieve $\tilde{U}(T_c) \to 1$, the primary goal becomes first order decoupling, that is, guaranteeing that at least $\overline{H}^{(0)} = 0$. To illustrate the method, consider the simplest possible system, that of a single spin-1/2 (qubit) in two situations: (a) $H_0 = B_z \sigma_z$, and (b) $H_0 = \vec{B} \cdot \vec{\sigma} = B_x \sigma_x + B_y \sigma_y + B_z \sigma_z$, where $\sigma_x, \sigma_y, \sigma_z$ are Pauli matrices, $B_z$ is a magnetic field in the $z$ direction, and $\vec{B}$ is a magnetic field in a supposedly unknown direction.

To freeze system (a) one needs to frequently undo the phase evolution by rotating the spin $180^o$ around a direction perpendicular to $z$. This may be accomplished with a sequence of $\pi$ pulses $P^x = \exp[-i\pi\sigma_x/2]$ applied after every $\Delta t$, as determined by the group $G = \{1, \sigma_x\}$. Cyclicity is ensuring by subjecting the system to an even number of pulses. Because the two transformed Hamiltonians, $\tilde{H}_1 = B_z\sigma_z$ and $\tilde{H}_2 = B_z\sigma_x\sigma_z\sigma_x = -B_z\sigma_z$, commute, exact cancellation of all orders $(k)$ in the average Hamiltonian is achieved at every $T_c$, leading to $\tilde{U}(nT_c) = U(nT_c) = 1$. Notice that this ideal result does not hold when dealing with realistic *finite* pulses, in which case $\overline{H} \neq 0$ and especial strategies have been developed to eliminate the first order terms in $\overline{H}$ [19, 67].

In general, however, even when the system is subjected to bang-bang pulses, the transformed Hamiltonians do not commute. This is the case of system (b). Here, the cancellation of $\overline{H}^{(0)}$ requires alternating $180°$ rotations of the spin around two perpendicular axes, so that each cycle consists of four pulses. An option corresponds to having $P_1 = P_3 = \exp[-i\pi\sigma_x/2]$ and $P_2 = P_4 = \exp[-i\pi\sigma_y/2]$, although any other path chosen to traverse the group $G = \{1, \sigma_x, \sigma_z, \sigma_y\}$ is viable. The four non-commuting transformed Hamiltonians, $\tilde{H}_1 = B_x\sigma_x + B_y\sigma_y + B_z\sigma_z$, $\tilde{H}_2 = B_x\sigma_x - B_y\sigma_y - B_z\sigma_z$, $\tilde{H}_3 = -B_x\sigma_x - B_y\sigma_y + B_z\sigma_z$, and $\tilde{H}_4 = -B_x\sigma_x + B_y\sigma_y - B_z\sigma_z$, lead to $\overline{H}^{(0)} = 0$, but $\overline{H}^{(1)} \neq 0$. A deterministic pulse sequence, which guarantees only first order decoupling, has been named periodic dynamical decoupling (PDD).

### 5.2. STRATEGIES TO IMPROVE PROTOCOL PERFORMANCE

The design of dynamical decoupling protocols aims at increasing averaging accuracy in the effective Hamiltonian and at slowing down the accumulation of residual averaging errors.

Here, we give an overview of some deterministic and randomized strategies to achieve these goals and discuss the advantages of combining both approaches.

5.2.1. DETERMINISTIC SCHEMES: Strategies exist to push beyond PDD and eliminate or reduce higher order terms in $\overline{H}$. Time symmetrization, for instance, corresponds to reversing the pulse sequence, so that at every $T_n = 2nT_c$, $\overline{H}^{(1)}$ and all odd terms in $\overline{H}$ are cancelled. In the example above, any of the possible PDD sequences, $\tilde{U}(T_c) = U_s.U_r.U_q.U_p$, with $p \in \{1,2,3,4\}$, $q \in \{1,2,3,4\} - \{p\}$, $r \in \{1,2,3,4\} - \{p,q\}$, and $s \in \{1,2,3,4\} - \{p,q,r\}$, leads to

$$\overline{H}^{(1)} = \frac{-i(\Delta t)^2}{2T_c} \{[\tilde{H}_s, \tilde{H}_r] + [\tilde{H}_q, \tilde{H}_p]\}, \quad (25)$$

where the above simplified form was obtained by using the equality

$$\tilde{H}_p + \tilde{H}_q + \tilde{H}_r + \tilde{H}_s = 0 . \quad (26)$$

It is straightforward to verify that the symmetric sequence $\tilde{U}(2T_c) = U_p.U_q.U_r.U_s.U_s.U_r.U_q.U_p$ leads to $\overline{H}^{(1)} = 0$.

Whenever the basic PDD sequence requires only 4 pulses, second order decoupling may also be achieved by swapping the elements in pairs of subsequent transformed Hamiltonians during the interval [8n∆t,4n∆t] so that

$$\tilde{U}(2T_c) = U_r.U_s.U_p.U_q.U_s.U_r.U_q.U_p \quad (27)$$

This last alternative may be further extended to achieve third order decoupling at every $T_n = 6nT_c$. Using Eq. (25), $\overline{H}^{(2)}$ may be written as

$$\overline{H}^{(2)} = \frac{-(\Delta t)^3}{6T_c} \left\{ [(2\tilde{H}_p + \tilde{H}_q),[\tilde{H}_p,\tilde{H}_q]] + [(2\tilde{H}_s + \tilde{H}_r),[\tilde{H}_s,\tilde{H}_r]] \right\}, \qquad (28)$$

One may verify that this term is suppressed with a supercycle scheme built up by arranging three $8\Delta t$-sequences as given below:

$$\tilde{U}(6T_c) = (U_q.U_s.U_r.U_p.U_s.U_q.U_p.U_r).(U_p.U_s.U_q.U_r.U_s.U_p.U_r.U_q).(U_r.U_s.U_p.U_q.U_s.U_r.U_q.U_p)$$

We refer to this protocol as the H2-scheme.

Concatenation [68, 69] and cyclic permutations [66] are other examples of strategies to improve protocol performance. Half of the concatenation procedure at second level coincides with Eq. (5.2) [66], whereas cyclic permutations are inspired by the MLEV decoupling sequence in NMR [140, 141]. In the particular case of a single spin-1/2 system, concatenation has so far proved to be the most efficient scheme [82-84], a fact associated with the irreducibility of the group considered [66]. This should be contrasted with the reducible group employed in subsection 5.4, where we find in increasing order of performance: time symmetrization, concatenation, cyclic permutations, and the H2 scheme [66].

Procedures such as concatenation and cyclic permutations do not necessarily cancel higher-order terms in $\overline{H}$, but they slow down their accumulation in time by varying the path to traverse the group. In periodically repeated sequences, the accumulation of residual errors caused by imperfect averaging is coherent (quadratic in time) and therefore extremely detrimental for long time evolutions. The key ingredient for efficient averaging at long times is to frequently scramble the order of the applied dynamical decoupling pulses, an idea which is at the heart of randomized methods.

5.2.2. RANDOMIZED SCHEMES: The most straightforward randomized dynamical decoupling protocol is obtained by picking elements uniformly at random over the decoupling group $G$, such that the control action at each $t_n = n\Delta t$ ($t_0 = 0$ included) corresponds to $P^{(r)} = g_i g_j^+$, $i, j = 0,1,...,|G|-1$. Such scheme is expected to outperform deterministic protocols at long times, but not at short times. High-level randomized protocols ensuring good performance at both short and long times are constructed by merging together advantageous deterministic and stochastic features. One option consists in selecting a deterministic sequence that guarantees high power of $\Delta t$ in the effective average Hamiltonian and embedding it with random pulses [85], which slows down error accumulation. Another alternative consists in randomly choosing at every $T_n = n|G|\Delta t$ a control path to traverse the group [88]; this sequence may be further improved if it is time symmetrized [87, 66].

A main characteristic of randomized methods is the great variety of control realizations; analyses of protocol performance are then based on averages over large samples of realizations, which are indicated in the figures below by $< >$. It is this enormous number of possible control realizations associated with large systems and long final times that hinders the search for optimal deterministic sequences at arbitrary times and favors randomization.

5.3. FRAMES AND CONTROL. The theoretical design of pulse sequences and the evaluation of their performances are usually done in the logical frame, whereas experiments are actually performed in the physical frame. These differences are disregarded when dealing with periodic sequences, since the two frames coincide at the end of each cycle, but acyclic sequences (as randomized schemes) may require a frame-correcting pulse before data acquisition [87, 66, 89].

In realistic control settings, in order to modify the system dynamics, the system is coupled, for instance, to an oscillating control field linearly polarized in the $x$ direction according to

$$H_c(t) = 2\Omega(t)\cos[\omega_f t + \varphi(t)]\frac{X}{2}, \qquad (29)$$

where $X, Y, Z$ correspond respectively to $\sigma_x, \sigma_y, \sigma_z$ in the case of a single spin-1/2 system and to $\sum_{i=1}^{N}\sigma_{x,i}$, $\sum_{i=1}^{N}\sigma_{y,i}$, $\sum_{i=1}^{N}\sigma_{z,i}$ for a system of $N$ spins-1/2. The experimentalist has control of the amplitude (power) $2\Omega$, the carrier frequency $\omega_f$, and the phase $\varphi$, as well as the interval $\tau$ during which $H_c(t)$ is on, and the separation $\Delta t$ between successive pulses.

All the results of this section are provided in a frame rotating with the frequency $\omega_f$ of the carrier. In this frame, by using the rotating wave approximation, the control Hamiltonian becomes

$$H_c^R(t) = \Omega(t)\left[\frac{X}{2}\cos\varphi(t) + \frac{Y}{2}\sin\varphi(t)\right]. \qquad (30)$$

The control field is applied in resonance with the frequency of the spin we want to rotate. The phase $\varphi$ determines the direction around which the rotation is realized in the rotating frame, and, in the case of rectangular pulses, $\Omega\tau$ characterizes the rotation angle. For example, a $\pi$ pulse around the $x$ direction requires $\Omega\tau = \pi$ and $\varphi(t) = 0$.

In the idealized scenario of bang-bang pulses, as considered here, the power is infinity and the pulse duration is zero. However, a complete analysis of dynamical decoupling protocols requires also the consideration of non-idealities such as finite pulses, flip-angle errors, and transients [19, 66], as well as pulse shapes [142-144].

5.4. ISOLATED HEISENBERG SPIN-1/2 SYSTEM WITH NEAREST-NEIGHBOR INTERACTIONS. Here, we show the advantages of randomization in dynamical decoupling methods applied for the case of a chain with $N$ spin-1/2 particles (qubits) coupled via isotropic nearest-neighbor interactions, as described by the Heisenberg model

$$H_0 = H_Z + H_{NN} = \sum_{i=1}^{N} \frac{\omega_i}{2} \sigma_{z,i} + \sum_{i=1}^{N-1} J \vec{\sigma}_i \cdot \vec{\sigma}_{i+1}, \qquad (31)$$

where $\omega_i$ is the Zeeman splitting enery of qubit $i$, $J$ is the coupling parameter between the spins, and open boundary conditions are assumed. This Hamiltonian models quasi-one-dimensional magnetic compounds [145] and Josephson-junction-arrays [146]. It is also a fairly good approximation for couplings which decay with the distance between the qubits – for cubic decay see [87].

Our goal is to freeze the evolution of the system for long times. We assume the possibility of individually addressing the spins with selective pulses and study the system in a combined logical-rotating frame, whereby one-body terms are removed from the Hamiltonian, so that $\tilde{H}_0^R \approx H_{NN}$. First order decoupling can be achieved through a very simple system-size-independent scheme. It consists of alternating two rotations around perpendicular axes, which act only on the odd qubits, or only on the even qubits, or yet the sequence has one direction associated with odd qubits and the other direction linked to even ones. The cycle is closed after four collective pulses. A possible representation of the control group for $N$ even is then given by

$$G = \{1, \sigma_{z,1}\sigma_{z,3}...\sigma_{z,N-1}, \sigma_{z,1}\sigma_{y,2}\sigma_{z,3}\sigma_{y,4}...\sigma_{z,N-1}\sigma_{y,N}, \sigma_{y,2}\sigma_{y,4}...\sigma_{y,N}\}. \qquad (32)$$

The path leading to the pulses $P_1 = P_3 = \exp\left(-i\pi\sum_{j=1,3}^{N-1}\sigma_{j,z}/2\right)$ and $P_2 = P_4 = \exp\left(-i\pi\sum_{j=2,4}^{N}\sigma_{j,y}/2\right)$, gives the four transformed Hamiltonians:

$$\tilde{H}_1 = \sum_{i=1}^{N-1} J\sigma_{x,i}\sigma_{x,i+1} + \sum_{i=1}^{N-1} J\sigma_{y,i}\sigma_{y,i+1} + \sum_{i=1}^{N-1} J\sigma_{z,i}\sigma_{z,i+1},$$

$$\tilde{H}_2 = -\sum_{i=1}^{N-1} J\sigma_{x,i}\sigma_{x,i+1} - \sum_{i=1}^{N-1} J\sigma_{y,i}\sigma_{y,i+1} + \sum_{i=1}^{N-1} J\sigma_{z,i}\sigma_{z,i+1}, \quad (33)$$

$$\tilde{H}_3 = \sum_{i=1}^{N-1} J\sigma_{x,i}\sigma_{x,i+1} - \sum_{i=1}^{N-1} J\sigma_{y,i}\sigma_{y,i+1} - \sum_{i=1}^{N-1} J\sigma_{z,i}\sigma_{z,i+1},$$

$$\tilde{H}_4 = -\sum_{i=1}^{N-1} J\sigma_{x,i}\sigma_{x,i+1} + \sum_{i=1}^{N-1} J\sigma_{y,i}\sigma_{y,i+1} - \sum_{i=1}^{N-1} J\sigma_{z,i}\sigma_{z,i+1}.$$

Among the deterministic strategies described in Sec.5.2.1, the supercycle H2 sequence

$$\tilde{U}(6T_c) = U_2.U_4.U_3.U_1.U_4.U_2.U_1.U_3.U_1.U_4.U_2.U_3.U_4.U_1.U_3.U_2.U_3.U_4.U_1.U_2.U_4.U_3.U_2.U_1, \quad (34)$$

is by far the best deterministic protocol, since it is the only one guaranteeing third order decoupling, that is, it eliminates $\bar{H}^{(o)}$, $\bar{H}^{(1)}$, and also $\bar{H}^{(2)}$. In Figure 6, we compare the performance of this efficient deterministic protocol with two randomized schemes: EH2 represents an H2 sequence embedded with random pulses characterized by products of $\pi$-rotations performed at arbitrarily selected spins around any of the three randomly chosen directions $x$, $y$, or $z$; and RH2 corresponds to a third order decoupling sequence where the path for the interval $[24n\Delta t, 24n\Delta t + 4\Delta t]$ is picked at random.

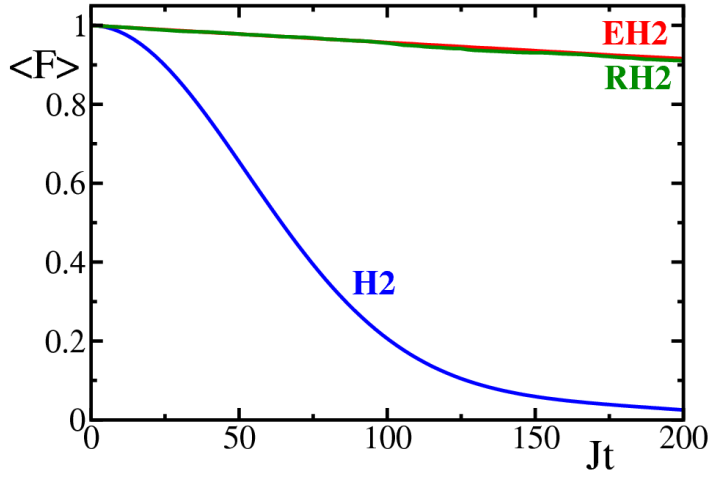

*Figure 6*: System described by $\tilde{H}_0^R \approx H_{NN}$ with $N=8$. Data acquired at $T_n = 24n\Delta t$, $\Delta t = 0.1 J^{-1}$. Blue line: deterministic H2 scheme leading to $\overline{H}^{(o)}, \overline{H}^{(1)}, \overline{H}^{(2)} = 0$; green line: randomized protocol RH2 constructed by randomly choosing a group path for each interval $[24n\Delta t, 24n\Delta t + 4\Delta t]$; red line: randomized protocol EH2 obtained by embedding the H2 sequence with random pulses. Average over 100 realizations.

The quantity considered to quantify protocol performance is the input-output fidelity (for other possibilities, see):

$$F(t) = |<\psi(0)|U(t)|\psi(0)>|^2, \qquad (35)$$

where we consider as initial pure state an eigenstate of a random matrix belonging to a Gaussian Orthogonal Ensemble.

At long times, the randomized protocols are significantly better, the fidelity decay being much slower than for the deterministic method. Both schemes, embedding the deterministic sequence with random pulses or applying path randomization, showed similar performance, but whether it is better to consider one or the other depends on the system at hand [87, 66, 88]. Even though an optimal deterministic sequence may exist for a particular system at a specific final time, identifying it may be very hard, in which case resorting to a simple, yet efficient randomized sequence, such as the one described here, is a more practical strategy.

### 5.5. SUPRESSION OF DECOHERENCE: SPIN-1/2 COUPLED TO A BOSONIC BATH

Consider a target system $S$ consisting of a spin-1/2 (qubit) coupled to a bosonic environment $E$ corresponding to independent harmonic modes, as described by the total Hamiltonian

$$H_0 = H_S + H_E + H_{SE}, \tag{36}$$

where

$$\begin{aligned} H_S &= \frac{\omega_0}{2}\sigma_z, \\ H_E &= \sum_k \omega_k b_k^+ b_k, \\ H_{SE} &= \sigma_z \sum_k g_k (b_k^+ + b_k), \end{aligned} \tag{37}$$

$\omega_0$ is the Zeeman splitting of the spin, $b_k^+$ and $b_k$ denote creation and annihilation bosonic operators of the environmental mode $k$ with frequency $\omega_k$, and $g_k$ determines the coupling parameter between the system and mode $k$. The system-bath coupling $H_{SE}$ leads to a purely dephasing process, spin population being unaffected by the environment. The advantage of such simple model is allowing for exact analytical derivations.

The purpose of dynamical decoupling here is to average out the evolution generated by $H_{SE}$ and prevent decoherence. After every $\Delta t$, the deterministic sequence corresponds to subjecting the system to a pulse $P^x = \exp[-i\pi\sigma_x/2]$, whereas for the randomized scheme, we choose at random between rotating or not the spin. Studies of protocol performance are based on the behavior of the off-diagonal elements of the reduced density matrix $\rho_{01}$, which is obtained after tracing over the degrees of freedom of the reservoir; $\rho_{01}$ contains all relevant phase information. The analysis is developed in a frame that removes both the control field and the free evolution due to $H_S$, which is referred to as logical-IP frame, IP standing for interaction picture.

By assuming that the system and the environment are initially uncorrelated and that the bath is in thermal equilibrium at temperature $T$ (the Boltzmann constant is set equal to 1), the expression for the system coherence in the logical-IP frame at time $t_n = n\Delta t$ is given by

$$\tilde{\rho}_{01}^I(t_n) = \rho_{01}(0)\exp[-\Gamma(t_n)], \tag{38}$$

where $\Gamma(t_n)$ is the decoherence function. For an ohmic bath in the continuum limit, we find:

In the absence of control,

$$\Gamma(t_n) = \alpha\int_0^\infty d\omega\, \omega e^{-\omega/\omega_c} \coth\left(\frac{\omega}{2T}\right)\frac{1-\cos[\omega t_n]}{\omega^2}. \tag{39}$$

For the deterministic scheme [76, 56],

$$\Gamma(t_n) = \alpha\int_0^\infty d\omega\, \omega e^{-\omega/\omega_c} \coth\left(\frac{\omega}{2T}\right)\frac{1-\cos[\omega t_n]}{\omega^2}\tan^2\left(\frac{\omega\Delta t}{2}\right). \tag{40}$$

For the randomized scheme [65],

$$\Gamma(t_n) = \alpha \int_0^\infty d\omega \omega e^{-\omega/\omega_c} \coth\left(\frac{\omega}{2T}\right) \frac{1-\cos[\omega t_n]}{\omega^2} \left[ n + 2\sum_{k=1}^{n-1} \cos(k\omega\Delta t) \sum_{l=0}^{n-k-1} \chi_l \chi_{l+k} \right], \quad (41)$$

where $\alpha$ is the interaction strength between the system and the bath, $\omega_c$ is an ultraviolet cutoff frequency, and $\chi_k$ is a Bernoulli random variable that accounts for the history of spin flips up to $t_k$ in a given realization, it takes the values $+1$ or $-1$ with equal probability [65].

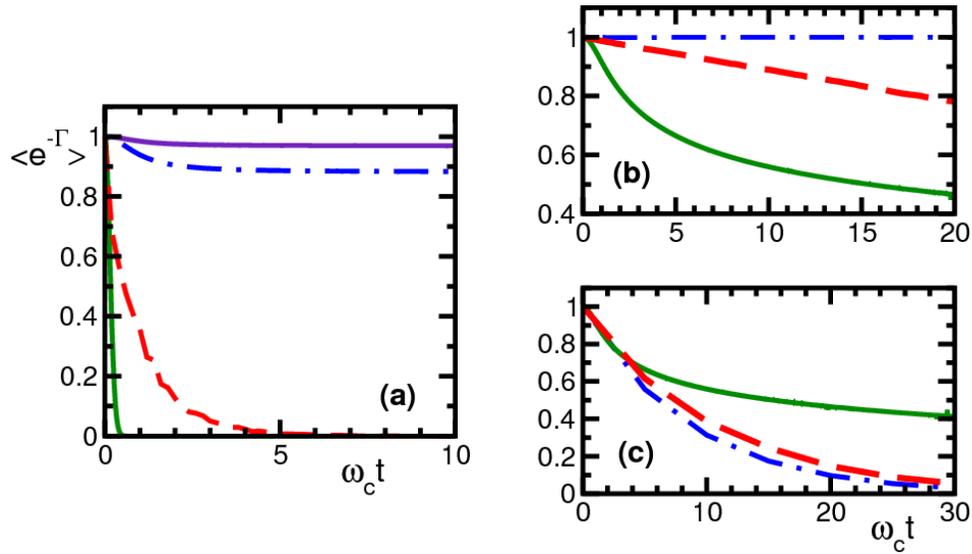

*Figure 7*: Decoherence rate from a bosonic ohmic bath. In units of $T^{-1}$: $\alpha = 0.25$, and $\omega_c = 100$. Left panel: $T = 100\omega_c$ and $\omega_c\Delta t = 0.1$. Right panels: $T = 0.01\omega_c$, top: $\omega_c\Delta t = 0.1$ and bottom: $\omega_c\Delta t = 2.5$. Green solid line: no control; red dashed line: randomized scheme; blue dot-dashed line: deterministic sequence; purple solid line (left panel): deterministic sequence with $\omega_c\Delta t = 0.05$. Average over 100 realizations.

Figure 7 compares the three decoherence functions above. Both high and low temperature baths are considered. Panel (a) shows the high temperature limit, which corresponds to an effectively classical bath dominated by thermal fluctuations. In the absence of control, decoherence is very fast on the time scale determined by the bath correlation time $\tau_c = \omega_c^{-1}$ and coherence preservation requires very short intervals between pulses. In the right panels, where the bath is at low temperature, decoherence is slower and a rich interplay between thermal and vacuum fluctuations occurs. Larger values of $\Delta t$ may be analyzed before total coherence loss takes place.

5.5.1. DECOHERENCE FREEZING: For short intervals between pulses, $\omega_c \Delta t \ll 1$, and at long times, $\omega_c t_n \gg 1$, it is seen from Eq. (40) that $\Gamma(t_n)$ becomes independent of $t_n$ being given by $\Gamma(t_n) = O(\alpha T \omega_c \Delta t^2)$ in the case of high temperatures, and by $\Gamma(t_n) = O(\alpha \omega_c^2 \Delta t^2)$ in the case of low temperatures. This asymptotic saturation is verified in panels (a) and (b) of Fig. 7, where $\omega_c \Delta t = 0.1$. The deterministic protocol eventually freezes decoherence at long times. This saturation was also verified in studies of an electron spin decohered by a nuclear spin environment in a quantum dot [82-84]. In NMR, saturation is associated to the "pedestals" seen in the long-time magnetization signal under pulsed spin-locking conditions [19].

5.5.2. LOW TEMPERATURE BATH AND DECOHERENCE ACCELERATION: The phenomenon of decoherence acceleration, where pulses induce destructive interference [56, 64], happens when the interval between pulses is larger than the correlation time of the bath, $\omega_c \Delta t > 1$. This is illustrated in panel (c). In experimental situations where the cutoff frequency of the reservoir cannot be overcome by the pulsing frequency, it is therefore better not to perturb the system. A similar scenario was encountered in Sec.2.1 where it was shown that tunneling becomes accelerated when the applied sequences of $2\pi$ pulses are not sufficiently frequent.

5.5.3. RANDOMIZATION AND STABILITY: In Fig. 7, whenever $\omega_c \Delta t < 1$, the randomized protocol is outperformed by the deterministic scheme. As a matter of fact, for the simple model of a single spin interacting with its environment, very efficient deterministic protocols have been identified for bosonic [72] as well as fermionic reservoirs [68, 69, 82-84].

Yet, the advantages of randomization in these models become perceptible when the system Hamiltonian is time dependent and little knowledge about it is available; randomization may then allow for enhanced stability against parameter variations [65]. In the case of a time-dependent system with more than one qubit, an illustration of the robustness of randomized schemes is provided in [87].

**ACKNOWLEDGMENTS**

L.G.C.R. gratefully acknowledges financial support from CNPq/Brazil. L.F.S. thanks Lorenza Viola for helpful discussions and start-up funds from Yeshiva University. V.S.B. acknowledges the NSF grants ECCS 0404191 and CHE 0345984, the DOE grant DE-FG02-07ER15909, and DOE supercomputer time from NERS